\documentclass{ws-p8-50x6-00}
\def\ANP{\em Adv. Nucl. Phys.}
\def\AP{\em Ann. Phys. (N.Y.)}
\def\NPA{\em Nucl. Phys. A}
\def\NC{\em Nuovo Cim.}
\def\NCA{\em Nuovo Cim. A}

\def\PA{\em Physica A}
\def\JPG{\em J. Phys. G}
\begin{document}

\title{Nucleon Properties in the Perturbative Chiral Quark Model} 

\author{\underline{V. E. Lyubovitskij}$^{\dagger, *}$, Th. Gutsche$^\dagger$ 
 and Amand Faessler$^\dagger$}
\address{$^\dagger$Institut f\"ur Theoretische Physik, Universit\"at 
T\"ubingen, Auf der Morgenstelle 14, D-72076 T\"ubingen, Germany,\\ 
$^*$Department of Physics, Tomsk State University, 634050 Tomsk, Russia, 
\\E-mail: Valeri.Lyubovitskij@uni-tuebingen.de,  
Thomas.Gutsche@uni-tuebingen.de, Amand.Faessler@uni-tuebingen.de}

\maketitle

\abstracts{We apply the perturbative chiral quark model (PCQM) to analyse
low-energy nucleon properties: electromagnetic form factors, meson-nucleon
sigma-terms and pion-nucleon scattering. Baryons are described as bound 
states of valence quarks surrounded by a cloud of Goldstone bosons 
($\pi, K, \eta$) as required by chiral symmetry. The model is based on the 
following guide lines: chiral symmetry constraints, fulfilment of low-energy 
theorems and proper treatment of sea-quarks, that is meson cloud 
contributions. Analytic expressions for nucleon observables are obtained 
in terms of fundamental parameters of low-energy pion-nucleon physics 
(weak pion decay constant, axial nucleon coupling constant, strong 
pion-nucleon form factor) and of only one model parameter (radius of the 
nucleonic three-quark core). Our results are in good agreement with 
experimental data and results of other theoretical approaches.}

\section{Introduction}

Hadron models set up to understand the structure of the nucleon should 
respect the constraints imposed by chiral symmetry. Spontaneous and explicit 
chiral symmetry breaking require the existence of the pion whose mass vanishes 
in the limit of zero current quark mass. In turn, nucleon observables must 
receive contributions from pion loops. 

We recently suggested \cite{PCQM} a baryon model, the perturbative chiral 
quark model (PCQM), which includes relativistic quark wave functions and 
confinement and respects chiral symmetry. The PCQM was successfully applied 
to $\sigma$-term physics \cite{PCQM} and extended to the study of 
electromagnetic properties of the nucleon \cite{PCQM1} and $\pi N$ 
scattering \cite{PCQM2}. Although the Lagrangian for this model fulfills the 
$SU(2) \times SU(2)$ current algebra, it is in general nontrivial to identify 
the explicit dynamics which leads to well-known predictions for various 
observables. For instance, $S$-wave $\pi N$ scattering at threshold is such 
a process which should be correctly described by a nucleon model involving 
the pion cloud. 

\section{Perturbative Chiral Quark Model}

\subsection{Effective Lagrangian and zeroth order properties}

Following considerations are based on the perturbative chiral quark model 
(PCQM), a relativistic quark model suggested in \cite{Gutsche} and extended 
in \cite{PCQM} for the study of low-energy properties of baryons. Similar 
models have also been studied in Refs.~\cite{Theberge-Chin}. 
The PCQM is based on the effective, chirally invariant Lagrangian 
${\cal L}_{inv}$ \cite{PCQM}
\begin{eqnarray}\label{Linv} 
{\cal L}_{inv}&=&\bar\psi \biggl\{ i\not\! \partial -\gamma^0 V(r)
-  \, S(r) \, \biggl[ \frac{U \, + U^\dagger}{2} \, 
+ \,\gamma^5 \, \frac{U \, -  U^\dagger}{2} \biggr] \biggr\}\,\psi\nonumber\\
&+&\frac{F^2}{4} \, {\rm Tr} [\, \partial_\mu U \, 
\partial^\mu U^\dagger \, ] 
\end{eqnarray}
where $\psi$ is the quark field, $U$ is the chiral field and $F=88$ MeV is the 
pion decay constant in the chiral limit \cite{PCQM,Gasser1}. The quarks move 
in a self-consistent field, represented by scalar $S(r)$ and vector $V(r)$ 
components of an effective static potential providing confinement. The 
interaction of quarks with Goldstone bosons is introduced on the basis of the 
nonlinear $\sigma$-model \cite{Gell-Mann_Levy}. We define the chiral field 
as $U=\exp[i\hat\Phi/F]$ using the {\it exponentional parametrization} of 
the nonlinear $\sigma$-model, where $\hat\Phi$ is the matrix of pseudoscalar 
mesons. When treating mesons fields as small fluctuations we do perturbation 
theory in the expansion parameter $1/F$. Usually, the Lagrangian (\ref{Linv}) 
is linearized with respect to the field $\hat\Phi$. Such an approximation is 
valid if we consider processes without free pions (e.g. nucleon mass shift 
due to the pion cloud) or with a single external pion field (e.g. pion-nucleon 
form factor). The resulting approximate chiral invariance of the linearized 
Lagrangian~\cite{PCQM,Thomas1} guarantees both a conserved axial current 
(or PCAC in the presence of the meson mass term) and the Goldberger-Treiman 
relation. However, when we study $\pi N$ scattering the quadratic term in the 
expansion of the chiral field $U$ should be kept. With the additional 
nonlinear term we reproduce~\cite{PCQM2} the model-independent result for the 
$\pi N$ $S$-wave scattering lengths as derived by derived by Weinberg 
and Tomozawa \cite{Weinberg_Tomozawa}. 

Here we concentrate on the application of the PCQM to the meson-nucleon 
$\sigma$-terms \cite{PCQM} and the electromagnetic properties of the 
nucleon \cite{PCQM1}. We use the effective Lagrangian 
${\cal L}_{eff} = {\cal L}_{inv}^{lin} + {\cal L}_{\chi SB}$  
including the linearized chiral-invariant term ${\cal L}_{inv}^{lin}$ and  
a mass term ${\cal L}_{\chi SB}$ which explicitly breaks chiral symmetry 
\begin{eqnarray}\label{Lagrangian_lin_inv}
{\cal L}_{inv}^{lin}&=&\bar\psi [i\not\!\partial - S(r) - \gamma^0V(r)]\psi + 
\frac{1}{2} (\partial_\mu\hat{\Phi})^2 
- \bar\psi S(r)i\gamma^5 \frac{\hat\Phi}{F} \psi , \nonumber\\   
{\cal L}_{\chi SB}&=&-\bar\psi {\cal M}\psi  -\frac{B}{8} 
{\rm Tr}\{ \hat\Phi , \, \{ \hat\Phi , \, {\cal M} \}\} ,  
\end{eqnarray}
where ${\cal M} = {\rm diag}\{\hat m, \hat m, m_s\}$ is the mass matrix of 
current quarks (here and in the following we restrict to the isospin symmetry 
limit with $m_u = m_d = \hat m$); B is the low-energy constant which measures 
the vacuum expectation value of the scalar quark densities in the chiral 
limit \cite{Gasser_Leutwyler}. We rely on the standard picture of chiral 
symmetry breaking \cite{Gasser_Leutwyler} and for the masses of pseudoscalar 
mesons we use the leading term in their chiral expansion (i.e. linear in the 
current quark mass):  $M_{\pi}^2=2 \hat m B, \hspace*{.2cm} 
M_{K}^2=(\hat m + m_s) B, \hspace*{.2cm} 
M_{\eta}^2= \frac{2}{3} (\hat m + 2m_s) B$. 
Meson masses satisfy the Gell-Mann-Oakes-Renner and the Gell-Mann-Okubo 
relation $3 M_{\eta}^2 + M_{\pi}^2 = 4 M_{K}^2$. In the evaluation we use the 
following set of QCD parameters: $\hat m$= 7 MeV,  
$m_s/\hat m$=25 and $B=M_{\pi^+}^2/(2\hat m)$=1.4 GeV. 
Introduction of the electromagnetic field $A_\mu$ into the PCQM is 
accomplished by standard minimal substitution into 
Eq. (\ref{Lagrangian_lin_inv}).  

To describe the properties of baryons which are modelled as bound states of 
valence quarks surrounded by a meson cloud we formulate perturbation theory. 
In our approach the mass (energy) $m_N^{core}$ of the three-quark core of 
the nucleon is related to the single quark energy ${\cal E}_0$ by 
$m_N^{core}=3\cdot {\cal E}_0$. 
For the unperturbed three-quark state we introduce the notation 
$|\phi_0>$ with the appropriate normalization $<\phi_0|\phi_0>=1$. 
The single quark ground state energy ${\cal E}_0$ and wave function (w.f.) 
$u_0(\vec{x})$ are obtained from the Dirac equation 
\begin{eqnarray}\label{Dirac_Eq}
[-i\vec{\alpha}\vec{\nabla}+\beta S(r)+V(r)-{\cal E}_0]u_0(\vec{x})=0 .
\end{eqnarray}    
The quark w.f. $u_0(\vec{x})$ belongs to the basis of potential eigenstates 
(including excited quark and antiquark solutions) used for expansion of the 
quark field operator $\psi(x)$. Here we restrict the expansion to the ground 
state contribution with $\psi(x)=b_0 u_0(\vec{x}) \exp(-i {\cal E}_0 t)$, 
where $b_0$ is the corresponding single quark annihilation operator. 
In Eq. (\ref{Dirac_Eq}) the current quark mass is not included to simplify 
our calculational technique. Instead we consider the quark mass term as a 
small perturbation. 

For a given form of the potentials $S(r)$ and $V(r)$ the Dirac 
equation (\ref{Dirac_Eq}) can be solved numerically. Here, for the sake of 
simplicity, we use a variational {\it Gaussian ansatz} \cite{PCQM} for the 
quark wave function given by the analytical form: 
\begin{eqnarray}\label{Gaussian_Ansatz} 
u_0(\vec{x}) \, = \, N \, \exp\biggl[-\frac{\vec{x}^2}{2R^2}\biggr] \, 
\left(
\begin{array}{c}
1\\
i \rho \, \frac{\displaystyle{\vec{\sigma}\vec{x}}}{\displaystyle{R}}\\
\end{array}
\right) , 
\end{eqnarray}      
where $N=[\pi^{3/2} R^3 (1+3\rho^2/2)]^{-1/2}$ is a constant fixed by the 
normalization condition $\int d^3x \, u^\dagger_0(x) \, u_0(x) \equiv 1$;  
$\chi_s$, $\chi_f$, $\chi_c$ are the spin, flavor and color quark wave 
functions, respectively. Our Gaussian ansatz contains two model parameters: 
the dimensional parameter $R$ and the dimensionless parameter $\rho$. 
The parameter $\rho$ can be related to the axial coupling constant $g_A$ 
calculated in zeroth-order (or 3q-core) approximation: 
\begin{eqnarray}\label{g_A}
g_A=\frac{5}{3} \biggl(1 - \frac{2\rho^2} {1+\frac{3}{2} \rho^2} \biggr) = 
\frac{5}{3} \frac{1+2\gamma}{3} ,  
\end{eqnarray}
where $\gamma = 9g_A/10 - 1/2$. The parameter $R$ can be physically 
understood as the mean radius of the three-quark core and is related to the 
charge radius $<r^2_E>^P_{LO}$ of the proton in the leading-order (LO) 
approximation as \cite{PCQM} 
\begin{eqnarray}\label{r2ep_LO}
<r^2_E>^P_{LO} \, = \, \frac{3R^2}{2} \, 
\frac{1 \, + \, \frac{5}{2} \, \rho^2}
{1 \, + \, \frac{3}{2} \, \rho^2} \, = \, 
R^2 \biggl( 2 - \frac{\gamma}{2} \biggr) . 
\end{eqnarray}
In our calculations we use the value $g_A$=1.25 as obtained in 
ChPT \cite{Gasser1}. We therefore have only one free parameter, $R$. In the 
numerical evaluation $R$ is varied in the region from 0.55 fm  to 0.65 fm. 

\subsection{Perturbation theory and nucleon mass}

Following the Gell-Mann and Low theorem we define the mass shift of the 
nucleonic three-quark ground state $\Delta m_N$ due to the interaction 
with Goldstone mesons as 
\begin{eqnarray}\label{Energy_shift}
\hspace*{-.25cm} 
\Delta m_N \, \doteq \, {^N\!\!<}\phi_0|  
\sum\limits_{n=1}^{\infty} \frac{i^n}{n!} \, 
\int \, i\delta(t_1) \, d^4x_1 \ldots d^4x_n \, 
T[{\cal L}_I(x_1) \ldots {\cal L}_I(x_n)] |\phi_0>_{c}^{N}    
\end{eqnarray} 
where ${\cal L}_I = - S(r) \bar\psi \, i \gamma^5  
(\hat\Phi/F) \, \psi$ is the interaction Lagrangian between mesons and quarks 
treated as a perturbation and subscript "c" refers to contributions from 
connected graphs only. We evaluate Eq. (\ref{Energy_shift}) at one loop with 
$o(1/F^2)$ using Wick's theorem and the appropriate propagators. For the quark 
field we use a Feynman propagator for a fermion in a binding potential. 
By restricting the summation over intermediate quark states to the ground 
state we get   
\begin{eqnarray} 
iG_\psi(x,y) &=& <\phi_0|T\{\psi(x)\bar\psi(y)\}|\phi_0> \\ 
&\to&u_0(\vec{x}) \bar u_0(\vec{y})\exp[-i{\cal E}_0(x_0-y_0)]\theta(x_0-y_0). 
\nonumber 
\end{eqnarray} 
For meson fields we use the free Feynman propagator for a boson field. 
Superscript $"N"$ in Eq. (\ref{Energy_shift}) indicates that the matrix 
elements are projected on the respective nucleon states. 

The total nucleon mass including one-loop corrections is given by 
\begin{eqnarray}\label{Baryon_mass}
\hspace*{-.3cm} 
m_N=3 ({\cal E}_0 + \gamma\hat{m}) + 
\sum\limits_{\Phi} d_N^\Phi \Pi (M_\Phi^2); 
\, d_N^\pi=\frac{171}{400}, 
d_N^K=\frac{6}{19}d_N^\pi, d_N^\eta=\frac{1}{57}d_N^\pi , 
\end{eqnarray} 
where $d_N^\Phi$ are the recoupling coefficients defining the partial 
contribution of the $\pi$, $K$ and $\eta$-meson cloud to the mass shift of 
the nucleon. The contribution of a finite current quark mass to the nucleon 
mass shift is taken into account perturbatively (for details see \cite{PCQM}). 
The self-energy operator $\Pi (M_\Phi^2)$ is given by 
\begin{eqnarray}\label{Sigma_Phi}
\Pi (M_\Phi^2) \, = \, - \, \biggl(\frac{g_A}{\pi F}\biggr)^2 \,\, 
\int\limits_0^\infty \frac{dp \, p^4}{w_\Phi^2(p^2)} \,\, F_{\pi NN}^2(p^2)   
\end{eqnarray} 
where $w_\Phi(p^2)=\sqrt{M_\Phi^2+p^2}$ is the meson energy with  
$p=|\vec{p}\, |$ and $F_{\pi NN}(p^2)$ is the $\pi NN$ form factor normalized 
to unity at zero recoil $(p^2=0)$:  
\begin{eqnarray}
F_{\pi NN}(p^2) = \exp\biggl(-\frac{p^2R^2}{4}\biggr) \biggl\{ 1 \, + \, 
\frac{p^2R^2}{8} \biggl(1 \, - \, \frac{5}{3g_A}\biggr)\biggr\} .  
\end{eqnarray}

\subsection{Meson nucleon sigma-terms and Feynman-Hellmann theorem}

The scalar density operators $S_i^{PCQM}$ $(i=u, d, s)$, relevant for the 
calculation of the meson-baryon sigma-terms in the PCQM, are defined as the  
partial derivatives of the model $\chi SB$ Hamiltonian 
${\cal H}_{\chi SB}=-{\cal L}_{\chi SB}$ with respect to the current 
quark mass of i-th flavor $m_i$: 
\begin{eqnarray}
S_i^{PCQM} \doteq \frac{\partial {\cal H}_{\chi SB}}{\partial m_i} 
\, = \, S_i^{val} \, + \, S_i^{sea} . 
\end{eqnarray} 
$S_i^{val}$ is the set of valence-quark operators coinciding with the 
ones obtained from the QCD Hamiltonian. The set of sea-quark operators 
$S_i^{sea}$ arises from the pseudoscalar meson mass term \cite{PCQM}.  
To calculate meson-baryon sigma-terms we perform the perturbative expansion 
for the matrix element of the scalar density operator $S_i^{PCQM}$ 
between unperturbed 3q-core states and then project it onto the respective 
baryon wave functions. 

As an example, the expression for $\sigma_{\pi N}$ is given by 
\begin{eqnarray}\label{Sigma_piN}
\sigma_{\pi N} = \hat m \, <p|S_u^{PCQM} \, + S_d^{PCQM}|p>  = 
3 \gamma\hat{m} \, + \, \sum\limits_{\Phi} d_N^\Phi \Gamma (M_\Phi^2)  
\end{eqnarray} 
where the first term of the right-hand side of Eq. (\ref{Sigma_piN}) 
corresponds to the valence quark, the second to the sea quark contribution. 
The vertex function $\Gamma (M_\Phi^2)$ is related to the partial derivative 
of the self-energy operator $\Pi (M_\Phi^2)$ with respect to $\hat{m}$: 
\begin{eqnarray}\label{Gamma_piN}
\Gamma (M_\Phi^2) \, = \, 
\hat{m} \, \frac{\partial}{\partial \hat{m}} \, \Pi (M_\Phi^2) . 
\end{eqnarray}   
Using Eqs. (\ref{Baryon_mass}), (\ref{Sigma_piN}) and (\ref{Gamma_piN}) 
we directly prove the Feynman-Hellmann theorem 
$\sigma_{\pi N} = \hat{m} \cdot \partial m_N/\partial \hat{m}$.  

\section{Results}

\subsection{Meson-nucleon sigma-terms}

We start our numerical analysis with the $\pi N$ sigma-term. First, we 
restrict to the $SU(2)$ flavor picture. Numerically, the contribution of the 
valence quarks to the $\pi N$ sigma-term is $13.1$ MeV, the contribution of 
sea quarks at order $o(M_\pi^2)$ is $66.9 \pm 5.7$ MeV. Here and in the 
following the error bars are due to variation of the range parameter $R$ 
of the quark wave function (\ref{Gaussian_Ansatz}) from 0.55 to 0.65 fm. 
Taking into account higher-order contributions of the sea quarks we have the 
following result for the $\pi N$ sigma-term $\sigma_{\pi N}^{\pi}$ 
(superscript $\pi$ refers to the SU(2) flavor picture) of 
\begin{eqnarray}
\sigma_{\pi N}^{\pi} = 43.3 \pm 4.4 \,  {\rm MeV}.
\end{eqnarray}
The contributions of kaon and $\eta$-meson loops, $\sigma_{\pi N}^{K}$ and 
$\sigma_{\pi N}^{\eta}$ (superscripts $K$ and $\eta$ refer to the respective 
meson cloud contribution) are significantly suppressed relative to the pion 
cloud and to the valence quark contributions due to the energy denominators in 
the structure integrals. The same conclusion regarding the suppression of $K$ 
and $\eta$-meson loops was obtained in the cloudy bag model~\cite{Stuckey}. 
Numerically, kaon and $\eta$-meson cloud contributions are 
$\sigma_{\pi N}^{K} = 1.7 \pm 0.4$ MeV and 
$\sigma_{\pi N}^{\eta} = 0.023 \pm 0.006$ MeV. 
For the $\pi N$ sigma-term we have the following final value:  
\begin{eqnarray}
\sigma_{\pi N} = \sum\limits_\Phi \sigma_{\pi N}^\Phi = 
45 \pm 5 \,  {\rm MeV}. 
\end{eqnarray}
Our result for the $\pi N$ sigma-term is in perfect agreement with the 
value of $\sigma_{\pi N} \simeq 45$ MeV deduced by Gasser, Leutwyler and 
Sainio~\cite{GLS} using dispersion-relation techniques and exploiting the 
chiral symmetry constraints. 

Next we discuss our prediction for the strangeness content of the nucleon 
$y_N$ which is defined in the PCQM as 
\begin{eqnarray}
y_N = \frac{2 <p|S_s^{PCQM}|p>}{ <p|S_u^{PCQM} \, + \, S_d^{PCQM}|p>} . 
\end{eqnarray}
The direct calculation of the strange-quark scalar density $<p|S_s^{PCQM}|p>$ 
is completely consistent with the indirect one applying the Feynman-Hellmann 
theorem $<p|S_s^{PCQM}|p> = \partial m_N/\partial m_s$. The small value of 
$y_N = 0.076 \pm 0.012$ in our model is due to the suppressed contributions 
of kaon and $\eta$-meson loops. Our prediction for $y_N$ is smaller than the 
value $y_N \simeq 0.2$ obtained in~\cite{GLS} from an analysis of 
experimental data on $\pi N$ phase shifts. On the other hand, our prediction 
is quite close to the result obtained in the cloudy bag model 
$y_N \approx 0.05$~\cite{Stuckey}.  

For the $KN$ sigma-terms we obtain
\begin{eqnarray}
\sigma_{KN}^u  = 340 \pm 37 \, {\rm  MeV} \hspace*{.2cm}  
\mbox{and} \hspace*{.2cm} \sigma_{KN}^d  =  284 \pm 37 \, {\rm MeV} , 
\end{eqnarray}
which within uncertainties is consistent with values deduced in 
HBChPT and lattice QCD (see discussion in Ref.~\cite{PCQM}).  
Hopefully, future DA$\Phi$NE experiments at Frascati 
will allow for a determination of the $KN$ sigma-terms and hence for a better 
knowledge of the strangeness content of the nucleon. 

\subsection{Electromagnetic properties of the nucleon}

We extended our model analysis to the description of basic electromagnetic 
properties of the nucleon. Formal details can be found in Ref.~\cite{PCQM1}. 
We start with results for the magnetic moments of nucleons, $\mu_p$ and 
$\mu_n$. For our set of parameters we obtain: 
\begin{eqnarray} 
\mu_p = 2.62 \pm 0.02, \hspace*{.25cm}
\mu_n = -2.02 \pm 0.02, \hspace*{.25cm} \mbox{and} \hspace*{.25cm} 
\frac{\mu_n}{\mu_p} = - 0.76 \pm 0.01 . 
\end{eqnarray}
The leading order (LO, three-quark core) $\mu_p^{LO}$, $\mu_n^{LO}$ and 
next-to-leading order (NLO, meson cloud and finite current quark mass) 
contributions $\mu_p^{NLO}$, $\mu_n^{NLO}$ to the magnetic moments are 
given by 
\begin{eqnarray} 
& &\mu_p^{LO} = 1.8 \pm 0.15 , \hspace*{.25cm}
\mu_n^{LO} \equiv - \frac{2}{3} \mu_p^{LO} ,\\
& &\mu_p^{NLO} = \mu_p - \mu_p^{LO} = 0.82 \pm 0.13, \hspace*{.25cm}
\mu_n^{NLO} = \mu_n - \mu_n^{LO} = - 0.82 \pm 0.08 . \nonumber 
\end{eqnarray}
For the electromagnetic nucleon radii we obtain  
\begin{eqnarray} 
& &r^p_E = 0.84 \pm 0.05 \, \mbox{fm}, \hspace*{.5cm}
<r^2>^n_E = - 0.036 \pm 0.003 \, \mbox{fm}^2, \\
& &r^p_M = 0.82 \pm 0.02 \, \mbox{fm}, 
\hspace*{.5cm} r^n_M = 0.85 \pm 0.01 \, \mbox{fm} . \nonumber  
\end{eqnarray}
The LO contributions to the charge radius of the proton 
(see Eq. (\ref{r2ep_LO})) and to the magnetic radii of proton and neutron are 
dominant  
\begin{eqnarray} 
r^{p; \, LO}_E \, = \, 0.77 \pm 0.06 \, \mbox{fm}, \hspace*{.25cm}
r^{p; \, LO}_M  \equiv  r^{n; \, LO}_M \, = \, 0.73 \pm 0.06 \, \mbox{fm} . 
\end{eqnarray}
For the neutron charge radius squared we get the observed (negative) sign, 
but its magnitude is smaller than the experimental value. As in the naive 
$SU(6)$ quark model, the LO contribution to the neutron charge radius is zero 
and only one-loop diagrams give nontrivial contributions to this quantity: 
\begin{eqnarray}
<r^2>^n_E = -0.036 \pm 0.003 \,\, {\rm fm}^2 . 
\end{eqnarray}
In Table 1 we summarize our results for the static electromagnetic properties 
of the nucleon in comparison to experimental data. 
Results on the $Q^2$-dependence of the electromagnetic form factors can be 
found in Ref. \cite{PCQM1}.  

\begin{table}[t]
\caption{\normalsize{Static nucleon properties.}}
\begin{center}
\begin{tabular}{|c|c|c|}
\hline
Quantity & Our Approach & Experiment \\
\hline
$\mu_{p}$ & 2.62 $\pm$ 0.02 & 2.793 \\
\hline
$\mu_{n}$ & -2.02 $\pm$ 0.02 & -1.913 \\
\hline
$\mu_n/\mu_p$ & -0.76 $\pm$ 0.01 & -0.68 \\
\hline
$r_E^p$ (fm) & 0.84 $\pm$ 0.05 & 0.86 $\pm$ 0.01 \\
\hline
$<r^2>^n_E$ (fm$^2$) & -0.036 $\pm$ 0.003 & -0.119 $\pm$ 0.004 \\
\hline
$r_M^p$ (fm) & 0.82 $\pm$ 0.02 & 0.86 $\pm$ 0.06 \\
\hline
$r_M^n$ (fm) & 0.85 $\pm$ 0.01 & 0.88 $\pm$ 0.07 \\
\hline
\end{tabular}
\end{center}
\end{table}

\section*{Acknowledgments}
This work was supported by the Deutsche Forschungsgemeinschaft 
(DFG, grant FA67/25-1). V.E.L. thanks the organizers for the invitation.

\end{document}